\documentclass{article}
\usepackage[dblblindworkshop,final]{neurips_2026}
\workshoptitle{AI-Native Academia: Authorship, Peer Review, and Conference Governance under AI}

\usepackage[utf8]{inputenc}
\usepackage[T1]{fontenc}
\usepackage{hyperref}
\usepackage{xurl}
\usepackage{booktabs}
\usepackage{tabularx}
\usepackage{array}
\usepackage{amsmath}
\usepackage{amssymb}
\usepackage{amsfonts}
\usepackage{microtype}
\usepackage{xcolor}
\usepackage{enumitem}
\usepackage{graphicx}
\usepackage{ragged2e}

\usepackage[table]{xcolor}

\usepackage{tcolorbox}
\tcbuselibrary{skins}

\newtcolorbox{claimbox}[1]{
    enhanced,
    colback=gray!5,           
    colframe=black!75,   
    leftrule=1.5mm,             
    rightrule=0mm,            
    toprule=0mm,              
    bottomrule=0mm,           
    arc=0mm,                  
    title={#1},               
    detach title,             
    before upper={\tcbtitle\par\smallskip}, 
    coltitle=black,   
    fonttitle=\fontfamily{ppl}\selectfont\bfseries,
    boxsep=1mm,
    left=2mm,
    right=2mm,
    top=1mm,
    bottom=1mm,
}

\newcolumntype{Y}{>{\RaggedRight\arraybackslash}X}
\newcolumntype{L}[1]{>{\RaggedRight\arraybackslash}p{#1}}
\setlist[itemize]{leftmargin=*,topsep=2pt,itemsep=1pt,parsep=0pt}
\setlist[enumerate]{leftmargin=*,topsep=2pt,itemsep=1pt,parsep=0pt}
\setlist[description]{leftmargin=2.5em,style=nextline,topsep=2pt,itemsep=2pt}

\title{AI-Native Academia Should Build a Community Verification Layer}
\author{Anonymous Authors}

\author{%
  Haotian Zong
  \\
  Department of Applied Mathematics and Statistics\\
  Johns Hopkins University\\
  Baltimore, MD, USA \\
  \texttt{hzong4@jh.edu} \\
  \And
  Luhao Zhang\\
  Department of Applied Mathematics and Statistics\\
  Johns Hopkins University\\
  Baltimore, MD, USA \\
  \texttt{luhao.zhang@jhu.edu}
}

\begin{document}
\maketitle

\begin{abstract}
This paper argues that AI-native academia should build a \emph{Community Verification Layer}: interoperable, policy-governed infrastructure that converts researchers' routine reading and checking into persistent, claim- and version-linked evidence, while routing scarce deeper verification toward uncertain, consequential, and neglected research. 
AI is lowering the cost of producing plausible research artifacts faster than the cost of accountable evaluation, yet much of the evaluation researchers already perform while reading, citing, implementing, or extending work is never made reusable. 
We propose a tiered ladder from comprehension notes to claim checks, audits, artifact execution, reproduction, replication, and downstream validation. 
The layer would elicit judgments before revealing social signals, preserve multidimensional disagreement, privately verify contributors while allowing flexible public disclosure, allocate attention under expertise and coverage constraints, and make verification labor citable, with credit tied to the quality and depth of the evidence.
We distinguish this design from prior open and post-publication review and outline governance requirements and potential counterarguments.
\end{abstract}

\section{Introduction}
\label{sec:introduction}

Suppose an AI-assisted paper presents a theorem and an empirical result. One researcher checks its key lemma, another executes the code, a reading group finds an incomparable baseline, and a later team identifies a narrower validity regime. Each has produced useful evidence, but it remains scattered across private notes, repositories, referee reports, and social media. Later readers must reconstruct the same context or substitute venue, author, institution, citations, and visibility for evidence about the claims themselves.
This coordination failure becomes more consequential as AI changes research production. 
For example, end-to-end systems can already generate ideas, code, experiments, and manuscripts in bounded ML settings \citep{lu2024aiscientist,yamada2025aiscientistv2}. 
Indeed, NeurIPS main-track submissions rose from 9,467 in 2020 to 21,575 in 2025, supported by more than 20,000 reviewers and extensive senior calibration \citep{neurips2025review}. 
Producing a paper, however, does not by itself add reliable knowledge.
Only after its claims have undergone appropriate review and verification can the work be provisionally added to the shared scientific record.
As AI accelerates paper production, preserving this filter places a rapidly growing burden on finite verification capacity.

\begin{claimbox}{Position}
AI-native academia should build an interoperable Community Verification Layer that turns researchers' routine reading into auditable, claim- and version-linked evidence and deliberately allocates scarce human verification toward uncertain, consequential, and neglected research.
\end{claimbox}

\section{The core problem: verification scarcity}
\label{sec:diagnosis}

AI lowers the marginal cost of producing \textit{plausible} research artifacts across ideation, synthesis, drafting, coding, experiment planning, and revision. 
AI can also assist with verification, and LLM feedback can overlap with useful human comments \citep{liang2024feedback}. 
However, current automated verification systems still miss many serious errors \citep{son2025spot,xi2025flaws,bastounis2026navier}.
The gap between useful assistance and dependable verification therefore remains substantial.
Moreover, even if frontier AI eventually becomes a much stronger verifier, accountable human involvement will remain necessary because scientific claims enter a shared record for which people and institutions must take responsibility.
Researchers and institutions must be able to understand and contest the evidence rather than delegate acceptance entirely to AI.
Together, the result is a \textit{production--verification asymmetry} and \textit{signal shrinkage}: 
(i) when new artifacts create verification demand faster than the community can process it, unresolved claims accumulate as \emph{verification debt};
(ii) as polished prose, figures, equations, and citations become cheaper to generate, presentation quality becomes a weaker proxy for reliability \citep{ma2026engineering,loven2026scarcities}, making verification even harder.

Verification capacity and need are also heterogeneous.
A paper may be understandable but wrong, correct but unimportant, or valuable but unreplicated.
Table~\ref{tab:layers} therefore separates three layers that a single rating would conflate.
A theorem, causal analysis, security claim, and systems artifact require different expertise, while contributors also differ in competence, conflicts, opportunity cost, and independence. The relevant questions are which claim to check, at what depth, by whom, and at what cost. The wider community need not understand every artifact directly, but it needs a traceable path from consequential claims to versions, evidence, and unresolved uncertainty.

\begin{table}[ht]
  \caption{Comprehension, verification, and valuation require different evidence.}
  \label{tab:layers}
  \centering
  \footnotesize
  \setlength{\tabcolsep}{4pt}
  \renewcommand{\arraystretch}{1.3}
  \begin{tabularx}{\linewidth}{@{} L{0.19\linewidth} Y Y @{}}
    \toprule
    \textbf{Layer} & \textbf{Central question} & \textbf{Suitable evidence} \\
    \midrule
    \rowcolor{black!3}
    \textbf{Comprehension}
      & What does the artifact claim, assume, and exclude?
      & Summaries, claim maps, scope and limitation annotations. \\
    \textbf{Verification}
      & Are claims supported by proofs, data, code, experiments, and citations?
      & Claim checks, audits, execution, reproduction, replication. \\
    \rowcolor{black!3}
    \textbf{Valuation}
      & Is the work useful, important, or relevant---and to whom?
      & Domain judgment, downstream use, extensions, field-specific assessment. \\
    \bottomrule
  \end{tabularx}
\end{table}

\section{Why current institutions are insufficient}
\label{sec:priorart}

\emph{Open and portable evaluation}. Early proposals already envisioned aggregating assessments from papers researchers read, multidimensional judgments, reviewer reputation, and incentives for neglected work \citep{florian2012aggregating,kriegeskorte2012open,kriegeskorte2012consensus}. Contemporary infrastructures implement public post-publication discussion \citep{pubpeer}, collaborative and curated preprint review \citep{prereview,sciety,pci}, portable pre-journal reports \citep{lemberger2019reviewcommons}, publish-then-review \citep{eisen2020publishthenreview}, modular records \citep{octopus}, and bounties \citep{researchhub}. Together, these systems make evaluations visible, durable, and reusable across prepublication and postpublication stages. A broader coverage objective remains missing: contributor-selected attention can leave coverage and evaluation depth uneven, while assigned workflows remain bounded to particular venues or programs.

\emph{Formal assignment within bounded review cycles}. Conference review and computational mechanisms assign reviewers under expertise, fairness, and manipulation constraints \citep{neurips2025review,stelmakh2021peerreview4all,jecmen2020manipulation,openreview}. These mechanisms organize evaluation around a submission cycle and a venue-level decision. After that decision, revisions, code releases, reproductions, replications, and downstream use often enter separate systems, leaving the review record and allocation process disconnected from later claim-level evidence and routine reading and checking. A continuing verification layer would carry forward both the evidence and the unresolved tasks that emerge at each stage.

A Community Verification Layer would connect durable evaluation records with active allocation of verification effort. It would label the depth of each check, attach evaluations to specific claims and versions, collect judgments before showing others' assessments, and make missing coverage visible. Recent work on shared verification infrastructure, machine-readable claims, claim--evidence interfaces, and tools that connect review comments to evidence provides compatible components \citep{loven2026scarcities,booeshaghi2026machinereadable,martinboyle2026papertrail,ghorbanpour2026peerispect}. Our contribution is to integrate these components into the institutional design summarized in Table~\ref{tab:comparison}, centered on persistent evidence capture and coverage-oriented allocation.

\begin{table}[ht]
  \caption{Functional comparison of existing system classes and the proposed layer.}
  \label{tab:comparison}
  \centering
  \footnotesize
  \setlength{\tabcolsep}{4pt}
  \renewcommand{\arraystretch}{1.3}
  \begin{tabularx}{\linewidth}{@{} L{0.30\linewidth}
      >{\Centering\arraybackslash}X
      >{\Centering\arraybackslash}X
      >{\Centering\arraybackslash}X @{}}
    \toprule
    \textbf{Dimension}
      & \textbf{Open / community}
      & \textbf{Conference review}
      & \textbf{Proposed layer} \\
    \midrule
    \rowcolor{black!3}
    \textbf{Captures open evaluation}
      & Yes & Limited & Yes \\
    \textbf{Active allocation}
      & Variable & Yes & Yes \\
    \rowcolor{black!3}
    \textbf{Verification depth}
      & Variable & One coarse task & Tiered \\
    \textbf{Anti-herding}
      & Rare & Sometimes & Blind-first \\
    \rowcolor{black!3}
    \textbf{Coverage objective}
      & Usually implicit & Submission-bound & Explicit \\
    \bottomrule
  \end{tabularx}
\end{table}

\section{A Community Verification Layer}
\label{sec:design}

The Community Verification Layer would complement journals, conferences, repositories, and research tools. As an interoperable layer, it would turn routine reading and checking into structured records tied to claims and artifact versions. Its design would distinguish verification depth, preserve independent judgments, protect contributors, and direct scarce effort toward uncertain, consequential, and neglected claims. The following subsections define its core objects, contribution workflow, allocation rule, and privacy model.

\subsection{Objects, provenance, and verification depth}

The layer attaches a structured \texttt{Evaluation} to a paper or other research artifact. Each evaluation identifies the exact \texttt{ArtifactVersion} checked and a \texttt{Claim} linked to the relevant text, figure, theorem, data, citation, or code. It records scope, depth, judgment, confidence, effort, independence, conflicts, and material AI use. An inspectable \texttt{EvidenceRecord} supplies supporting evidence. Links to a privately verified contributor and later revisions, corrections, replications, or retractions preserve provenance: what was checked, against which version, with what evidence, and how the record changed. Persistent, versioned evaluations can also be cited as research outputs.
Nanopublications, micropublications, RO-Crate, and recent machine-readable claim--evidence interfaces show that evidence records of various levels can be structured and exchanged \citep{groth2010nanopublication,clark2014micropublications,soilandreyes2022rocrate,martinboyle2026papertrail}.
A tiered ladder prevents many shallow endorsements from outweighing one detailed audit. Table~\ref{tab:ladder} classifies contribution depth, where each level states what the contribution can support.

\begin{table}[ht]
  \caption{Proposed verification ladder.}
  \label{tab:ladder}
  \centering
  \footnotesize
  \setlength{\tabcolsep}{4pt}
  \renewcommand{\arraystretch}{1.3}
  \begin{tabularx}{\linewidth}{@{} L{0.27\linewidth} Y Y @{}}
    \toprule
    \textbf{Level} & \textbf{What is checked} & \textbf{Supported statement} \\
    \midrule
    \rowcolor{black!3}
    \textbf{0. Relevance note}
      & Topic, audience, relation to other work
      & \emph{``Relevant to X.''} \\
    \textbf{1. Comprehension}
      & Claims, assumptions, scope, limits
      & \emph{``This is what the artifact argues.''} \\
    \rowcolor{black!3}
    \textbf{2. Claim check}
      & One theorem step, citation, figure, or assumption
      & \emph{``I checked this component.''} \\
    \textbf{3. Technical audit}
      & Substantial proof, method, statistical, or security audit
      & \emph{``I audited these parts under these conditions.''} \\
    \rowcolor{black!3}
    \textbf{4. Artifact execution}
      & Running released code or inspecting supplied data
      & \emph{``The supplied artifact produces these outputs.''} \\
    \textbf{5. Reproduction}
      & Recreating results with substantially the same data and procedure
      & \emph{``The reported analysis can/cannot be reproduced.''} \\
    \rowcolor{black!3}
    \textbf{6. Independent replication}
      & Testing with independently generated evidence
      & \emph{``The finding does/does not generalize here.''} \\
    \textbf{7. Downstream validation}
      & Successful use, extension, deployment, or theoretical incorporation
      & \emph{``The result proved useful or valid in this context.''} \\
    \bottomrule
  \end{tabularx}
\end{table}

\subsection{Low-friction capture and blind-first judgment}

Researchers already evaluate papers when they cite, teach, check proofs, implement methods, or compare baselines, but most of this work stays private. The layer should capture it in forms that match the effort spent: a short note may mark relevance or scope, while a deeper contribution may document a concern, test, or reproduced result. Required structure should grow with claimed depth and fit common research tools. 
Records should capture a multidimensional profile of claim support, method or proof validity, reproducibility, scope, importance, and confidence.
The interface should show judgment distributions and evidence depth, including negative and inconclusive checks that reveal unresolved uncertainty and verification cost.

Contributors should record an initial assessment before seeing aggregate judgments. They may then inspect judgment distributions, author responses, and requests for deeper work. 
The system should also keep initial and revised reports separate. 
Experiments in broader domains show that visible judgments from others can change assessments, increase inequality, or reduce diversity without improving accuracy \citep{muchnik2013social,salganik2006inequality,lorenz2011wisdom}. Blind-first elicitation is therefore a precaution to preserve independent information. 

\subsection{Active allocation of scarce verification}

Passive systems record what researchers happen to review. The Community Verification Layer should instead rank contributor--claim--depth pairings by the expected value of additional evidence. One illustrative priority score is
\begin{equation}
P_{ij\ell}\propto
\frac{\mathrm{Fit}_{ij}\,\mathrm{Uncertainty}_{i}\,\mathrm{Value}_{i}\,\mathrm{Neglectedness}_{i}\,\mathrm{Independence}_{ij}\,\mathrm{MarginalGain}_{i\ell}}
{\mathrm{ExpectedCost}_{ij\ell}},
\label{eq:allocation}
\end{equation}
where $i$ is a claim or task, $j$ is a contributor, and $\ell$ is a verification depth. 
The score provides a transparent starting point for prioritization, and pilot deployments would estimate each term and test how well the resulting rankings predict realized verification gains and costs.
$\mathrm{Fit}$ is task competence; $\mathrm{Uncertainty}$ is unresolved evidence; $\mathrm{Value}$ reflects the consequences of truth or error; $\mathrm{Neglectedness}$ rewards coverage gaps; $\mathrm{Independence}$ discounts shared labs, models, code, and networks; $\mathrm{MarginalGain}$ estimates what depth $\ell$ would add for claim $i$; and $\mathrm{ExpectedCost}$ includes time, access, compute, and opportunity cost. Thus, a first reproduction may add more than another summary.
The layer could flag an unchecked theorem, a disputed assumption, or code that has run but not been reproduced. Assignment research offers tools for fair assignment and resistance to manipulation \citep{stelmakh2021peerreview4all,jecmen2020manipulation}. The layer would adapt these tools to maintain a continuing evidence record and direct follow-up verification across claims and versions.

\subsection{Accountable privacy}

Accountability can rest on private identity verification and enforceable governance while allowing contributors to control public disclosure. Evidence on reviewer identity is mixed: open-review research raises concerns about power dynamics, a randomized trial found small benefits from keeping reviewers anonymous to one another during discussion, and an ICLR incident report documents risks following identity exposure \citep{rosshellauer2024scoping,rastogi2024anonymizing,iclr2026retrospective}. 
The consortium administering the Community Verification Layer should link each contribution to a verified person and enforce the rules, where contributors could verify identity through institutional credentials, persistent identifiers, or another trusted process, and then choose public signing, a stable pseudonym, identity escrow, or delayed disclosure.

\section{Alternative views}
\label{sec:counterarguments}

\paragraph{One ladder cannot fit all sciences.}
Proof checking, causal inference, qualitative interpretation, benchmark audits, and systems reproduction use different standards. The ladder provides a shared structure for contribution type and depth, while communities should define field-specific evidence profiles. Across fields, the schema would record target, version, scope, evidence, credential, conflicts, provenance, confidence, and update history.

\paragraph{Verification will legitimize low-quality artifacts.}

An open verification layer could also be flooded with plausible but low-value artifacts, and even ``unverified'' may confer legitimacy. 
The layer should therefore separate basic record admission from expert-attention allocation. 
Triage should check basic completeness, deter mass creation, account for scarce expert time, and display ``no qualified evaluation recorded.'' 
It would fail if it ends up rewarding volume rather than evidence.

\section{Conclusion}
\label{sec:conclusion}

AI-native academia needs infrastructure that makes verification legible, reusable, deliberately allocated, and contestable over time. A Community Verification Layer would preserve otherwise-private understanding, distinguish evidence depths, expose neglected claims, and route deeper work under fairness, independence, privacy, and governance constraints. 

\bibliographystyle{plain}
\bibliography{ref}

\end{document}